\documentclass[journal=chreay,manuscript=article]{achemso}
\usepackage{graphicx}
\usepackage{amsmath}
\usepackage{multirow}
\usepackage{textcomp}
\usepackage{hyperref}
\usepackage[font=small,labelfont=bf]{caption}
\usepackage{setspace}
\title{
On the Balance of Intercalation and Conversion Reactions in Battery Cathodes
}

\author{Daniel C. Hannah}
\affiliation{
Materials Science Division, Lawrence Berkeley National Laboratory,
Berkeley, CA 94720, USA}
\alsoaffiliation{These authors contributed equally to this work}

\author{Gopalakrishnan Sai Gautam}
\affiliation{
Department of Materials Science and Engineering, Massachusetts
Institute of Technology, Cambridge, MA 02139, USA}
\alsoaffiliation{
Materials Science Division, Lawrence Berkeley National Laboratory,
Berkeley, CA 94720, USA}
\alsoaffiliation{
Department of Materials Science and Engineering, University of California Berkeley,
CA 94720, USA}
\alsoaffiliation{These authors contributed equally to this work}

\author{Pieremanuele Canepa}
\affiliation{
Materials Science Division, Lawrence Berkeley National Laboratory,
Berkeley, CA 94720, USA}
\alsoaffiliation{
Department of Materials Science and Engineering, Massachusetts
Institute of Technology, Cambridge, MA 02139, USA}

\author{Gerbrand Ceder} \email{gceder@berkeley.edu, gceder@lbl.gov}
\affiliation{
Department of Materials Science and Engineering, University of California Berkeley,
CA 94720, USA}
\alsoaffiliation{
Materials Science Division, Lawrence Berkeley National Laboratory,
Berkeley, CA 94720, USA}
\alsoaffiliation{
Department of Materials Science and Engineering, Massachusetts
Institute of Technology, Cambridge, MA 02139, USA}

\begin{document}


\begin{abstract}
We present a thermodynamic analysis of the driving forces for intercalation and conversion reactions in battery cathodes across a range of possible working ion, transition metal, and anion chemistries.  Using this body of results, we analyze the importance of polymorph selection as well as chemical composition on the ability of a host cathode to support intercalation reactions.  We find that the accessibility of high energy charged polymorphs in oxides generally leads to larger intercalation voltages favoring intercalation reactions, whereas sulfides and selenides tend to favor conversion reactions.  Furthermore, we observe that Cr-containing cathodes favor intercalation more strongly than those with other transition metals. Finally, we conclude that two-electron reduction of transition metals (as is possible with the intercalation of a $2+$ ion) will favor conversion reactions in the compositions we studied.
\end{abstract}


\section{Introduction}
\label{sec:intro}
Multivalent (MV) batteries, such as those based on Mg, Ca, and Zn, can potentially offer substantial gains in volumetric energy density via non-dendritic stripping and deposition of a metal anode.\cite{whittingham2014,aurbach2001study, ponrouch2015towards, liu2016dendrite,aurbach2000prototype,canepa2017odyssey,guo2015_ang} Furthermore, the intercalation of divalent working ions may potentially be combined with multi-redox transition metals, enabling high-capacity cathodes.\cite{canepa2017odyssey}  To date, the poor mobility of MV ions in most solid frameworks constitutes a major obstacle to their utilization in practical intercalation batteries.{\cite{canepa2017odyssey}} Previous studies have indicated that using cathode hosts with an ``un-preferred'' coordination environment can mitigate poor MV mobility.{\cite{rong2015materials}} Since cathode structures without the MV ions present, i.e., ``charged'' cathodes such as V$_2$O$_5$, MoO$_3$ and MnO$_2$,{\cite{imhof1999,amatucci2001_v2o5,gershinsky2013electrochemical,Xu2013_Zn,bruce1991}} are more likely to intercalate the MV ions into an un-preferred coordination, most MV electrochemical experiments have attempted MV intercalation into a charged host structure, unlike Li-ion and Na-ion systems.{\cite{whittingham2014,komaba2014_nareview,guo2017_aem}} Additionally, most cathodes in modern Li-ion batteries follow an intercalation pathway during electrochemical discharge, while MV battery cathodes have been shown in a number of cases to undergo conversion reactions,\cite{canepa2017odyssey,arthur2014understanding, zhang2016unveil, pan2016reversible, cawo3conversion,zhang2015_jps} wherein a reduced transition metal oxide is formed alongside a thermodynamically stable alkaline earth metal oxide, such as MgO. In Mg systems, the occurrence of conversion reactions is thought to be driven predominantly by the stability of MgO, which creates a large thermodynamic driving force for conversion.\cite{canepa2017odyssey,ling2015general,wang2015_chemical} Because conversion reactions of the type described above are reduction reactions, they occur only upon battery discharge, when the cathode material is being reduced.\par

In commercial batteries, the (de)intercalation of ions into cathodes, such as Li in layered-Li$_{\rm x}$CoO$_2$,{\cite{mizushima1980lixcoo2}} occurs without altering the crystal structure of the host material, a process commonly denoted as ``topotactic'' (de)intercalation. A recent thermodyanmic study by Ling \emph{et al.}\cite{ling2015general} finds that topotactic Mg insertion into K-$\alpha$MnO$_2$ is thermodynamcally unfavorable compared to conversion into MgO and various Mn oxide binaries, consistent with experimental efforts.\cite{arthur2014understanding} In their report, Ling and co-workers suggest that a propensity toward conversion (rather than intercalation) may be a general phenomenon in Mg battery cathodes. Indeed, at least one other report has directly examined this phenomenon and demonstrated a preference for conversion in polyanion systems such as olivine-FePO$_4$.\cite{zhang2016unveil} However, the occurrence of conversion reactions can depend strongly on polymorph selection for a given cathode chemistry. For example, while K-$\alpha$MnO$_2$ was conclusively shown to convert to MgO and MnO,\cite{arthur2014understanding, ling2015general} $\lambda$-MnO$_2$ appears to exhibit some degree of reversible Mg insertion.\cite{canepa2017odyssey, sinha2008electrochemical, cabello2015electrochemical, kim2015direct} Notably, $\alpha$-MnO$_2$ (charged composition) and $\lambda$-MgMn$_2$O$_4$ (Mg-discharged composition) can both be experimentally synthesized,{\cite{vicat1986,Irani1962}} highlighting the importance of the starting structure in favoring conversion or (de)intercalation.\par

For most cathode hosts, the energetic balance between conversion and intercalation remains unknown, particularly for less-studied working ions such as Ca$^{2+}$ and Zn$^{2+}$. Additionally, there are several as-yet unexplained occurrences of capacity fade in MV cathode materials,\cite{sun2016layered, kaveevivitchai2016high, zhang2012alpha, orikasa2014high, feng2008preparation} which may be attributable to conversion reactions.\par

In this work, we establish a thermodynamic framework for the examination and comparison of conversion and intercalation energetics in cathode hosts across the chemical and structural spaces.  As an example, we consider the Mg-Cr-$X$ ($X$~=~O, S, Se) ternary systems to decipher the primary thermodynamic driving forces for intercalation or conversion, including polymorph selection and anion chemistry. Finally, we leverage high-throughput, first-principles calculations to examine the balance of conversion and intercalation for five important working ions (Li, Na, Ca, Mg, Zn) within a comprehensive set of transition metal oxide, sulfide, and selenide hosts. \par

Our findings demonstrate that most oxide materials favor intercalation provided the extent of discharge is limited to one-electron reduction of the transition metal. On the other hand, similar estimates suggest favorable conversion reactions for sulfide and selenide chemistries. For two-electron reduction of the transition metal, our results indicate that conversion reactions are always thermodynamically favored. Importantly, we find that cathode frameworks that are the lowest energy forms at the discharged composition (e.g., $\lambda$-MgMn$_2$O$_4$) are more resistant to conversion reactions than the lowest energy structures at the corresponding charged composition ($\alpha$-MnO$_2$), in agreement with existing experimental findings.{\cite{arthur2014understanding,Xu2013_Zn,kim2015direct,nam2015high,sun2016investigation}} Interestingly, we find that Cr-containing compounds are the most resistant to conversion, regardless of working ion or anion, which we attribute to an ideal balance of Cr and anion chemical potential in the Cr-chalcogenide binaries. Finally, we emphasize the importance of metastability and kinetic stabilization, particularly in multivalent cathodes, where the ability of oxides to attain highly metastable configurations can facilitate topotactic intercalation at high voltages, while certain kinetically stabilized compounds (such as MgTi$_2$S$_4$) are known to exhibit reversible intercalation despite a (small) thermodynamic driving force for conversion.\cite{sun2016high}

\section{Thermodynamics of intercalation and conversion}
\label{sec:framework}
When considering the reduction of a host framework by a working ion, either an intercalation reaction or a conversion reaction can occur. These reactions can be generally described as: 

\begin{center}
${\rm A}^{m+} + ze^- + n{\rm MX}_2 \rightarrow {\rm A}({\rm MX}_{2})_n$ (Intercalation) \\
${\rm A}^{m+} + ze^- + n{\rm MX}_2 \rightarrow \sum_i {\rm A}_p^{\left(i\right)}{\rm M}_q^{\left(i\right)}{\rm X}_r^{\left(i\right)}$ (Conversion)
\end{center}

\noindent where $A$ stands for the working ion (Li, Na, Ca, Mg, or Zn), ``M'' is a 3$d$ transition metal, and $X$ is an anion (O, S, or Se). The sum in the conversion reaction runs over the set of all reaction products \{$i$\}. Charge balance requires that $n = m/z$, with $z$ and $m$ being the number of electrons involved in the reduction of the cathode host and the valence of the working ion, respectively. Thus, the reaction A~+~MX$_2 \rightarrow$~A(MX$_2$), i.e. $n$~=~1, can either represent a 1 electron reduction of M by a monovalent ion or a 2 electron reduction by a divalent ion. For example, the magnesiation of MnO$_2$ might proceed as either of the reactions shown below:

\begin{center}
(Intercalation) MgMnO$_2 \longleftarrow$ Mg$^{2+}$ + 2$e^{-}$ + MnO$_2 \longrightarrow$ MgO + MnO~(Conversion)
\end{center}

\noindent The Gibbs free energies driving the conversion and intercalation reactions directly relate  to conversion ($V_{\rm conv}$)  and intercalation  ($V_{\rm int}$) voltages via $V = -\Delta G/zF$, as in Eq.~{\ref{eq:1}} and {\ref{eq:2}}, respectively.

\begin{equation}\label{eq:1}
V_{\rm int} = - \frac{G_{{\rm A}({\rm MX}_2)n} - G_{\rm A} - nG_{\rm MX_2}}{zF}
\end{equation}
\begin{equation}\label{eq:2}
V_{\rm conv} = - \frac{\sum_i G_i^{\rm conv} - G_{\rm A} - nG_{\rm MX_2}}{zF}
\end{equation}

\noindent where the Gibbs free energies $G$ are approximated by the 0 K enthalpies obtained from DFT calculations, i.e. $G \approx E_{\rm DFT}(0 \rm K)$, which explicitly ignores vibrational and configurational entropy contributions. Previous studies have shown that the voltage predictions obtained via DFT often benchmark well with experimental values.\cite{wang2006oxidation, zhou2004first, urban2016computational}.

The difference between the intercalation voltage ($V_{\rm int}$) and the conversion voltage ($V_{\rm conv}$) determines which reaction is favored, with a larger voltage difference implying a stronger thermodynamic driving force. In the present analysis we consider which reaction is more likely to occur upon electrochemical discharge ---i.e. beginning from high voltage. Thus, the higher voltage process (intercalation vs.\ conversion) will be the one that is likely to occur upon discharge.

\subsection{Classification of possible discharge voltages}
\begin{figure}[!h]
\begin{center}
  \includegraphics[width=0.85\columnwidth]{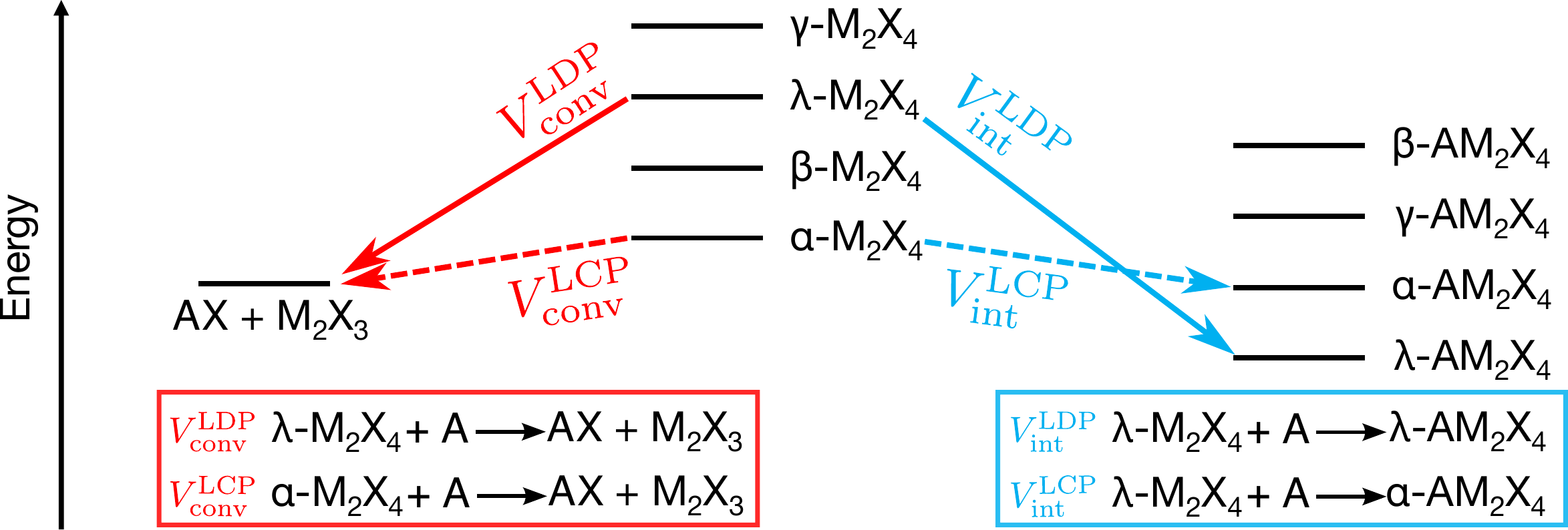}
  \end{center}
  \caption{\label{fig:1}
	(Color online) Summary of the structure selection scheme for the voltage calculations described in the manuscript. A is the working ion (Li, Na, Mg, Ca, or Zn), M is the $3d$ transition metal, and X is the anion (O, S, or Se). $\gamma$, $\lambda$, $\beta$, $\alpha$ refer to different polymorphs for M$_2$X$_4$. Energy refers to the Gibbs energy of the different polymorphs considered. For each chemistry, we calculate two voltages for both intercalation ($V_{\rm int}$, blue arrows) and conversion ($V_{\rm conv}$, red arrows), considering the lowest energy discharged polymorph ($V^{\rm LDP}$, solid arrows) and the lowest energy charged polymorph ($V^{\rm LCP}$, dashed arrows). The explicit intercalation and conversion reactions used for the voltage calculations are indicated in the highlighted blue and red boxes, respectively.}  
\end{figure}
While the higher voltage amongst intercalation ($V_{\rm int}$) and conversion ($V_{\rm conv}$) indicates the thermodynamically favored discharge process, different polymorphs can result in different intercalation and conversion voltages for a given cathode chemistry. In terms of electrochemical cycling of a cathode, there are two broad approaches used during synthesis: ($i$) the cathode is made at the discharged composition (i.e., the working ion is already contained within the cathode), as is common in Li-ion and Na-ion systems{\cite{mizushima1980lixcoo2,whittingham2004,delmas1981,kang2012_na_review}} or ($ii$) the cathode is at the charged composition (without the working ion), as is practiced in MV chemistries.{\cite{canepa2017odyssey,Levi2004,Levi2007,bruce1991}} Note that whenever a cathode is synthesized at the discharged (or charged) composition, the discharged (charged) polymorph with the lowest Gibbs energy is often obtained,{\cite{gibbs1878}} though some level of metastability is possible in synthesis.{\cite{sun2016thermodynamic,sun2017_nitrides}} Also, in most cathode chemistries, the lowest energy polymorphs at the discharged and charged compositions are significantly different.{\cite{arroyo2009gaining,yamada2006room,reed2004role,van1998first}} Thus, depending on whether the synthesized cathode has the same structure as the lowest energy discharged (charged) polymorph, the intercalation and conversion voltages obtained will differ, necessitating the calculation of four distinct voltages, as defined in the text below and summarized in Figure~{\ref{fig:1}} and Table~{\ref{table:1}}.

Given a charged M$_2$X$_4$ cathode which has the same structure as the lowest energy discharged polymorph (LDP), i.e., $\lambda$-M$_2$X$_4$ in Figure~{\ref{fig:1}} (solid arrows in Figure~{\ref{fig:1}}), there are two possible reactions: ($i$) topotactic intercalation of A into $\lambda$-M$_2$X$_4$ host to form $\lambda$-AM$_2$X$_4$, which is the lowest energy discharged polymorph. The voltage at which the intercalation reaction occurs is termed $V_{\rm int}^{\rm LDP}$ (solid blue arrow in Figure~{\ref{fig:1}}). ($ii$) conversion of the $\lambda$-M$_2$X$_4$ host upon reduction with A to form a combination of stable phases (e.g., AX+M$_2$X$_3$), where the conversion reaction occurs at $V_{\rm conv}^{\rm LDP}$ (solid red arrow). Analogously, starting with a charged-M$_2$X$_4$ host with the structure of the lowest energy charged polymorph (LCP), i.e., $\alpha$-M$_2$X$_4$ (dashed arrows in Figure~{\ref{fig:1}}), the cathode can either undergo intercalation to form $\alpha$-AM$_2$X$_4$ at $V_{\rm int}^{\rm LCP}$ (dashed blue arrow in Figure~{\ref{fig:1}}) or conversion to form a combination of stable phases at $V_{\rm conv}^{\rm LCP}$ (dashed red arrow).

Note that we calculate the potential of conversion reactions by assuming a decomposition of the cathode host to the nearest stable phases on the A-M-X ternary phase diagram upon reduction. For example, intercalated spinel-MgMn$_2$S$_4$ is thermodynamically unstable,{\cite{liu2016evaluation}} and its composition is bounded by MnS, MgS, and MnS$_2$ on the Mg-Mn-S phase diagram (see Figure~S1a in Supporting Information -- SI). Therefore, we consider the conversion reaction to be Mg + spinel-Mn$_2$S$_4 \rightarrow$ MnS + MgS + MnS$_2$. Similar analysis is also performed for systems with thermodynamically stable intercalated products. Even if an intercalation product is stable, we compute a hypothetical conversion voltage assuming decomposition to the neighboring (stable) phases on the ternary phase diagram. For example, the stable spinel-MgMn$_2$O$_4$ is bound by Mg$_6$MnO$_8$, Mn$_2$O$_3$ and Mn$_3$O$_4$ on the Mg-Mn-O phase diagram (see Figure~S1b), leading to the (hypothetical) conversion reaction 6~Mg + 6~(spinel-Mn$_2$O$_4) \rightarrow$ Mg$_6$MnO$_8$ + 4 Mn$_2$O$_3$ + Mn$_3$O$_4$.

\begin{table}
 \caption{\label{table:1} Voltage types calculated in this work.}
\begin{tabular*}{\columnwidth}{@{\extracolsep{\fill}}llll@{}}
  \hline\hline
 Voltage type & Nomenclature &Structure of cathode \\ 
 \hline
 \multirow{2}{*}{Intercalation} & $V_{\rm int}^{\rm LDP}$ & Lowest energy discharged polymorph\\ 

  &$V_{\rm int}^{\rm LCP}$ & Lowest energy charged polymorph \\

 \multirow{2}{*}{Conversion} & $V_{\rm conv}^{\rm LDP}$ & Lowest energy discharged polymorph\\ 

 & $V_{\rm conv}^{\rm LCP}$ & Lowest energy charged polymorph \\ 
 \hline \hline
\end{tabular*}
\end{table}

\subsection{Structure selection}
\label{sec:structure}
\noindent {\bf Composition --} Our analysis covers the compositional space of A(MX$_2$)$_n$ compounds, as described earlier in Section~{\ref{sec:framework}}. For a given cathode material A(MX$_2$)$_n$, we consider all structures from a crystal structure database{\cite{jain2013commentary,icsd}} matching the stoichiometry of the charged and discharged compostions. We choose A(MX$_2$)$_n$ because a plurality of known cathode materials are of this form, such as, LiCoO$_2$,\cite{mizushima1980lixcoo2} Li(MnO$_2$)$_2$ or LiMn$_2$O$_4$,\cite{thackeray1984electrochemical} MgMn$_2$O$_4$,\cite{kim2015direct} MgTi$_2$S$_4$,\cite{sun2016high} NaMnO$_2$,\cite{ma2011electrochemical} etc. \\

\noindent {\bf Metastability --} Because intercalation and conversion reactions can explicitly depend on the polymorph for a given cathode and working ion combination, a selection scheme is needed to identify which structures are considered. For a particular chemistry, the polymorphs used in typical battery cathodes are rarely the ground state at all working ion concentrations.\cite{van1998first, reed2004role, zhou2006configurational}  A prominent example is LiFePO$_4$ cathode material, in which electrochemical charging of the LiFePO$_4$ yields FePO$_4$ in the (thermodynamically unstable) olivine crystal structure despite berlinite being the most stable crystal structure at this composition.\cite{arroyo2009gaining, yamada2006room, malik2011kinetics, zhou2006configurational} Thus, for a topotactic charge/discharge process, at least one end member (charged or discharged) is always metastable. Thus for LCP (LDP) structures, the corresponding discharged (charged) polymorph is likely metastable.

A useful metric to quantify the extent of instability of a given structure is the energy above hull (E$^{\rm hull}$), which describes the amount of energy released by the decomposition of a compound into the most stable compounds at that composition. For example, spinel-MgMn$_2$S$_4$ is metastable and has an E$^{\rm hull} = 73$~meV/atom on the Mg-Mn-S ternary phase diagram, while the stable spinel-MgMn$_2$O$_4$ has an E$^{\rm hull} = 0$~meV/atom on the Mg-Mn-O ternary.{\cite{jain2013commentary}} While the usual guideline for synthesizability of a compound has been suggested to be E$^{\rm hull} < $~50 meV/atom,{\cite{liu2015spinel, liu2016evaluation, sun2016thermodynamic}} the true upper limits of metastability are not rigorously known and electrochemical cycling frequently yields metastable structures inaccessible to conventional synthesis methods. Therefore, to calculate the LDP (LCP) intercalation voltages, we consider the topotactically matched charged (discharged) structure irrespective of its instability (given by the magnitude of E$^{\rm hull}$).  \\

There are two caveats to the scheme detailed above:
\begin{enumerate}
	\item For compositions with unknown structures, we rely on prototype structures, analogous to those observed in Li- and Na-ion chemistries at each composition.{\cite{whittingham2014,komaba2014_nareview,kang2012_na_review}} For A(MX$_2$)$_2$ or AM$_2$X$_4$ compounds, the spinel structure is used (comparable to spinel-LiMn$_2$O$_4$.{\cite{thackeray1984electrochemical}}).  For AMX$_2$ compounds, a layered structure is used (analogous to layered-LiCoO$_2${\cite{mizushima1980lixcoo2}} and Na$_{\rm x}$CrO$_2${\cite{bo2016layered}}).  As an example, we used a layered structure as prototype for LiTiSe$_2$ since the actual structure is unknown. The specific compositions for which hypothetical structures are utilized are detailed in the SI (see Sections~S4, S6 and the \texttt{structures\_SI.xlsx} supplementary file).
	\item For LCP structures that do not have a corresponding topotactic discharged structure, we use a hypothetical discharged structure with a E$^{\rm hull} =$~100~meV/atom for the calculation of $V_{\rm int}^{\rm LCP}$. For example, rutile is the LCP of Cr$_2$O$_4$ but reliable structures of rutile-MgCr$_2$O$_4$ are unavailable. Therefore, to estimate a ``reasonable'' upper-bound for $V_{\rm int}^{\rm LCP}$, we used a  E$^{\rm hull} =$~100~meV/atom for rutile-MgCr$_2$O$_4$.  The specific compositions for which this approximation has been made are provided in the SI (\texttt{structures\_SI.xlsx}). Although a conversion reaction will be thermodynamically favored when the intercalated structure is highly unstable, using a maximum E$^{\rm hull} =$~100~meV/atom gives a useful estimate on the magnitude of the driving force for conversion. Note that the calculation of LDP voltages does not require any approximation on E$^{\rm hull}$ since the topotactic charged structure can always be (theoretically) obtained by the removal of the working ions from the LDP structure.
\end{enumerate}

\section{Computational methods}
We obtain optimized structures and internal energies from Density Functional Theory (DFT)\cite{kohn1965self} calculations as implemented in the Vienna \emph{ab-initio} Simulation Package (VASP).\cite{kresse1993ab} The exchange-correlation functional is approximated by the Perdew-Burke-Ernzerhoff (PBE) implementation of the Generalized Gradient Approximation (GGA).\cite{perdew1996generalized} The wavefunctions are described using the Projector Augmented Wave (PAW) theory\cite{kresse1996efficient} combined with a kinetic energy cutoff of 520 eV and are sampled on a Monkhorst-Pack mesh with a \emph{k}-point density of at least 1000/(number of atoms in the unit cell). When warranted, spurious self-interaction errors on $d$-electrons are accounted for by adding a Hubbard-$U$ correction. The $U$ values used in this work, which are listed in Section~S9 of the SI, were fitted to reproduce experimental transition metal oxidation enthalpies, as detailed originally by Jain \emph{et al.}\cite{jain2011formation}  The 0~K phase diagrams utilized in this work are constructed via the methods implemented in the pymatgen Python API\cite{ong2013python} and using the Materials Project\cite{jain2013commentary} database and are supplemented by local DFT calculations wherever needed (carried out according to the procedure described above). To be consistent with the DFT-calculation scheme implemented in Materials Project, our calculations include spin polarization, while we did not explicitly account for van der Waals interactions.

\section{Results}

All intercalation and conversion reactions used to derive the voltages discussed in this manuscript can be found in Sections~S3 and S5 of the SI.

\subsection{The Mg-Cr-X system}
\begin{figure}[h]
\begin{center}
  \includegraphics[width=\columnwidth]{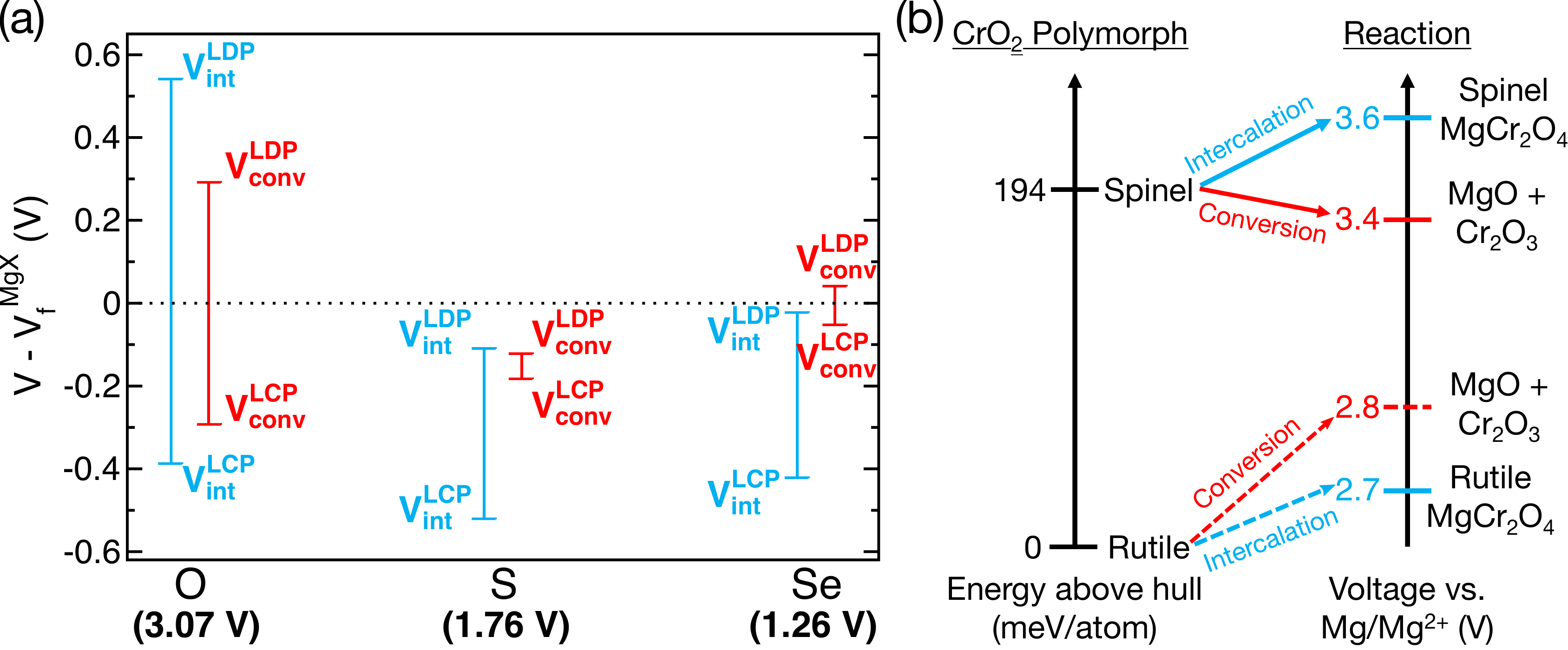}
  \end{center}
  \caption{\label{fig:2}
	(a) Range of possible intercalation (blue) and conversion voltages (red) for Mg intercalation into or reduction of Cr$_2$X$_4$ (X = O, S, Se) by Mg. The voltages are referenced to the voltage at which the binary MgX compound forms, i.e., voltage for the reaction Mg~+~X~$\rightarrow$~MgX, which is indicated in parenthesis below the corresponding anion along the $x$-axis. The polymorphs involved in the reactions are indicated in Table~{\ref{table:new}}. LDP and LCP indicate the lowest energy discharged and charged polymorphs, respectively. (b) A schematic of intercalation and conversion reactions for the LDP (spinel) and LCP (rutile) in the CrO$_2$ system. The left diagram in (b) displays the energetics of charged-CrO$_2$ polymorphs while the right diagram displays possible reduction products as a function of voltage vs. Mg/Mg$^{2+}$.}
\end{figure}

We first apply our intercalation vs.~conversion analysis for a 1$e^-$ (per transition metal ion) reduction in the Mg-Cr-X (X = O, S, Se) system, which is of particular interest in light of recent predictions that Cr$_2$O$_4$ and Cr$_2$S$_4$ spinels are promising cathode materials for MV batteries.\cite{chen2017_mgcr2o4,liu2015spinel, liu2016evaluation} Figure \ref{fig:2}a displays the intercalation (blue) and conversion (red) voltages for Mg discharge into Cr$_2$X$_4$, computed according to the scheme in Figure \ref{fig:1} and Table \ref{table:1}. To facilitate direct comparison across the anion chemistries, the voltages are referenced to the voltage of formation of the corresponding binary MgX, i.e., the voltage of the reaction Mg~+~X~$\rightarrow$~MgX, which is indicated along the $x$-axis beneath the corresponding anion. The LDP and LCP structures associated with each voltage and the actual voltages are indicated in Table~{\ref{table:new}}. For example, the LDP for all three anions is the spinel. Therefore the calculated $V^{\rm LDP}$ for both intercalation and conversion incorporates the energetics of the spinel structure for the charged state of Cr$_2$X$_4$. While the LCP of CrO$_2$ is rutile, the LCP of CrS$_2$ and CrSe$_2$ is a layered structure (Table~{\ref{table:new}}). Thus, the LDP and LCP reactions for Mg-discharge in the CrO$_2$ system, which are displayed in Figure~{\ref{fig:2}}b, can be summarized as:

\begin{center}
Mg + Cr$_2$O$_4$ (rutile) $\rightarrow$ MgCr$_2$O$_4$ (rutile) (LCP, intercalation) \\
Mg + Cr$_2$O$_4$ (rutile) $\rightarrow$ MgO + Cr$_2$O$_3$ (LCP, conversion) \\
Mg + Cr$_2$O$_4$ (spinel) $\rightarrow$ MgCr$_2$O$_4$ (spinel) (LDP, intercalation) \\
Mg + Cr$_2$O$_4$ (spinel) $\rightarrow$ MgO + Cr$_2$O$_3$ (LDP, conversion)
\end{center}

The LDP always has higher intercalation voltage than the LCP polymorph (Figure~{\ref{fig:2}}), which reflects the stability of the discharged state. The voltage is the energy lowering of the charged polymorph as the working ion is inserted, which is by definition larger for discharge from a metastable to a stable compound than the other way around. The data in Figure \ref{fig:2}a clearly demonstrates that the accessibility of high energy charged polymorphs in Cr$_2$O$_4$ creates a much wider intercalation voltage range ($\sim$~0.9~V, height of the blue bar in Figure~{\ref{fig:2}}a) than for Cr$_2$S$_4$ or Cr$_2$Se$_4$ ($\sim$~0.4~V). The intercalation voltage for the LDP structure (blue tick labeled $V_{\rm int}^{\rm LDP}$ in Figure~{\ref{fig:2}}a), is substantially higher than the corresponding conversion voltage (V$_{\rm conv}^{\rm LDP}$) in the oxide as compared to the sulfide or selenide, indicating a reduced tendency for conversion in the oxide. Although the larger voltage of MgCr$_2$O$_4$ suggests a more favorable energy density for oxides, it is worth noting that sulfides generally exhibit superior mobility for MV ions.{\cite{liu2015spinel, liu2016evaluation}} Note that for all Cr-X chemistries considered, Mg reduction of the LCP structure is expected to undergo conversion, as indicated by the higher $V_{\rm conv}^{\rm LCP}$ than $V_{\rm int}^{\rm LCP}$. There are important consequences to this finding as it may oppose the general desire to start from charged polymorphs for optimizing the mobility in multivalent systems.\par

While Figure \ref{fig:2}a compares the effects of anion chemistry, Figure \ref{fig:2}b emphasizes the role of polymorphism within (Mg)Cr$_2$O$_4$. Specifically, in Figure \ref{fig:2}b, we analyze the reaction of Mg with two charged-Cr$_2$O$_4$ polymorphs: rutile, which is stable, and spinel, which is 194 meV/atom above the compositional hull. In Figure \ref{fig:2}b, red and blue arrows indicate conversion and intercalation reactions, respectively. When reacting with rutile CrO$_2$, Mg preferably forms conversion products, MgO + Cr$_2$O$_3$, at 2.8~V, instead of the (hypothetical) intercalation product rutile MgCr$_2$O$_4$, which forms at a lower 2.7~V. However, a similar reaction of Mg with spinel CrO$_2$ proceeds differently, where the intercalated MgCr$_2$O$_4$ preferably forms at a higher 3.6~V compared to the conversion products (MgO + Cr$_2$O$_3$) at 3.4~V. The effect of polymorphism in the Mg-Cr-O system is qualitatively similar to the behavior experimentally observed in the Mg-Mn-O system, as discussed in Section~{\ref{sec:intro}}. Thus, the polymorph with which Mg discharge occurs can play a critical role in whether reversible intercalation (discharge into spinel Cr$_2$O$_4$) or irreversible conversion (rutile Cr$_2$O$_4$) occurs. Also, Table~{\ref{table:new}} indicates that for the Cr sulfide and selenide the conversion and intercalation voltage for the spinel (LDP) are very similar, signifying that if one could intercalate Mg at near-equilibrium conditions, the driving force for conversion would be small.

\begin{table}
 \caption{\label{table:new} All reactions considered for the Mg-Cr-O, Mg-Cr-S and Mg-Cr-Se systems and the associated voltages (in V), labeled according to the conventions adopted and described in Figure \ref{fig:1}.}
\begin{tabular*}{\columnwidth}{@{\extracolsep{\fill}}lcccr@{}}
\hline\hline
 Reaction & LDP structure & $V^{\rm LDP}$ & LCP structure & $V^{\rm LCP}$ \\
\hline
\multicolumn{5}{c}{\textbf{Mg-Cr-O} } \\
Mg + Cr(IV)$_2$O$_4$ $\rightarrow$ MgCr(III)$_2$O$_4$  & \multirow{2}{*}{Spinel} & 3.61 &  \multirow{2}{*}{Rutile} & 2.68 \\
Mg + Cr$_2$O$_4$ $\rightarrow$ MgO + Cr$_2$O$_3$  & & 3.36 &  & 2.78  \\
\hline
\multicolumn{5}{c}{\textbf{Mg-Cr-S}  } \\
Mg + Cr$_2$S$_4$ $\rightarrow$ MgCr$_2$S$_4$  & \multirow{2}{*}{Spinel} & 1.65 &  \multirow{2}{*}{Layered} & 1.24 \\
Mg + Cr$_2$S$_4$ $\rightarrow$ MgS + Cr$_2$S$_3$  & & 1.64 & & 1.58 \\
\hline
\multicolumn{5}{c}{\textbf{Mg-Cr-Se}  } \\
Mg + Cr$_2$Se$_4$ $\rightarrow$ MgCr$_2$Se$_4$  & \multirow{2}{*}{Spinel} & 1.28 &  \multirow{2}{*}{Layered} & 0.84 \\
Mg + Cr$_2$Se$_4$ $\rightarrow$ MgSe + Cr$_2$Se$_3$  & & 1.30 & & 1.21 \\
\hline
\hline
\end{tabular*}
\end{table}

\subsection{One-electron Reduction}

\begin{figure}[!h]
\begin{center}
  \includegraphics[width=\columnwidth]{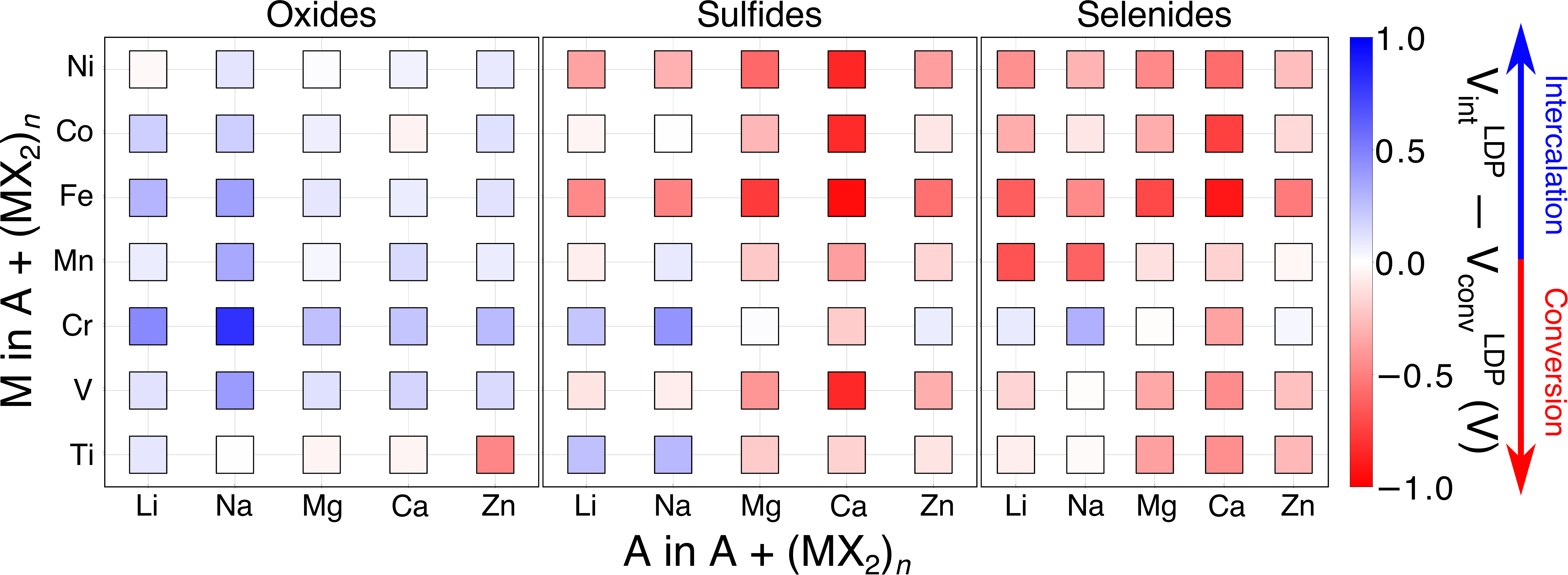}
  \end{center}
  \caption{\label{fig:3}
	Difference between the intercalation ($V_{\rm int}^{\rm LDP}$) and conversion ($V_{\rm conv}^{\rm LDP}$) voltage for 1-electron reduction reactions, starting from the lowest energy discharged polymorph (LDP). The voltage difference is indicated for five working ions (A~=~Li, Na, Mg, Ca, and Zn) in various $3d$-transition metal oxide, sulfide, and selenide hosts. Higher intercalation voltages are indicated by blue-colored squares while higher conversion voltages are red-colored. Note that the higher voltage indicates the thermodynamically favorable process. For monovalent ions (Li, Na), the intercalation is into a MX$_2$ structure, while for divalent ions it is into a M$_2$X$_4$ structure, corresponding to a 1$e^-$ (per transition metal ion) reduction.}
\end{figure}

\begin{figure}[!h]
\begin{center}
  \includegraphics[width=\columnwidth]{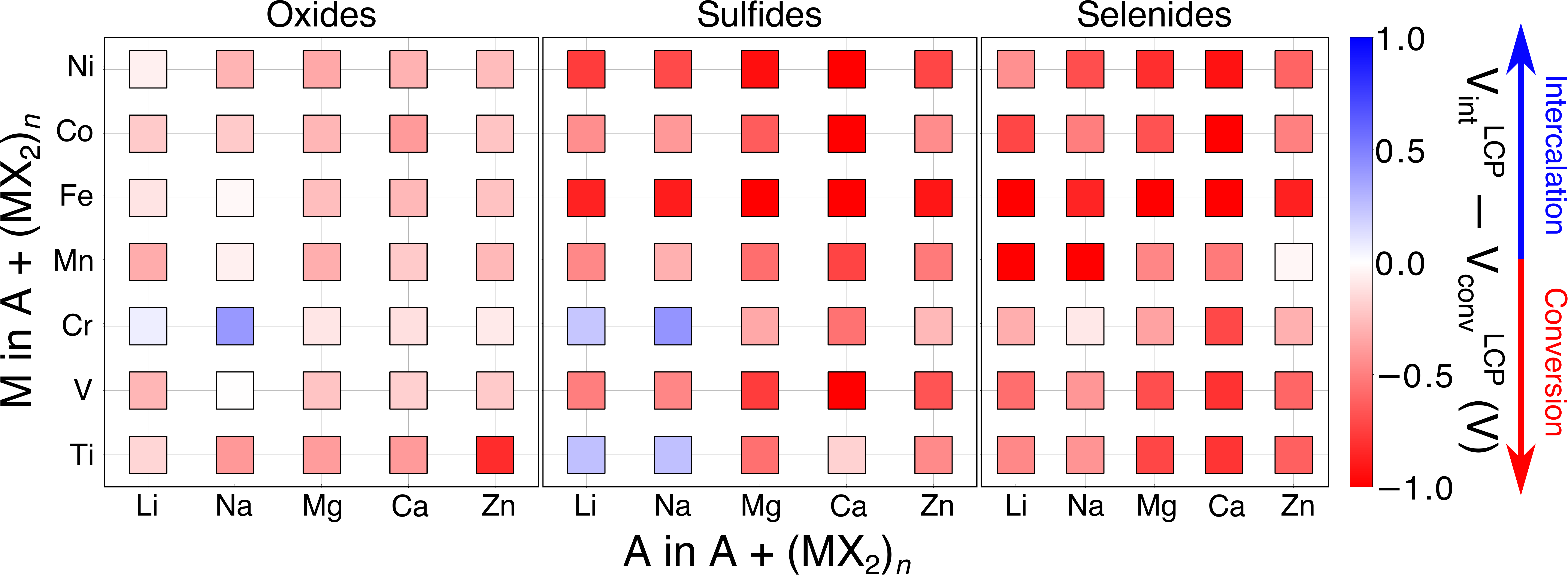}
  \end{center}
  \caption{\label{fig:3b}
	Difference between the intercalation ($V_{\rm int}^{\rm LCP}$) and conversion ($V_{\rm conv}^{\rm LCP}$) voltage for 1-electron reduction reactions, starting from the lowest energy charged polymorph (LCP). The voltage difference is indicated for five working ions (A~=~Li, Na, Mg, Ca, and Zn) in various $3d$-transition metal oxide, sulfide, and selenide hosts. Higher intercalation voltages are indicated by blue-colored squares while higher conversion voltages are red-colored. For monovalent ions (Li, Na), the intercalation is into a MX$_2$ structure, while for divalent ions it is into a M$_2$X$_4$ structure, corresponding to a 1$e^-$ (per transition metal ion) reduction.}
\end{figure}

Figures~{\ref{fig:3}} and {\ref{fig:3b}} indicate the difference between the intercalation and conversion voltage for a 1$e^-$ reduction process in $3d$-transition metal chalcogenide hosts, using the structure of the lowest energy discharged polymorph (LDP, Figure~{\ref{fig:3}}), and the lowest energy charged polymorph (LCP, Figure~{\ref{fig:3b}}), respectively. Calculated intercalation and conversion voltages are always referenced to the bulk metallic form of the working ion (Li voltages are against Li/Li$^{+}$, for example). Figures~\ref{fig:3} and {\ref{fig:3b}} is divided into panels representing an anion chemistry, including oxides (left panel), sulfides (center), and selenides (right). Each panel of Figures~{\ref{fig:3}} and {\ref{fig:3b}} consider the discharge of five working ions (A~=~Li, Na, Mg, Ca, and Zn) as indicated along the $x$-axis with each row on the $y$-axis corresponding to a $3d$-transition metal. While higher intercalation voltages for each combination of working ion, $3d$-metal, and anion is indicated by blue-colored squares, higher conversion voltages are indicated by red-colored squares. Note that the intercalation process considered for monovalent ions (Li, Na) is into a MX$_2$ host while for multivalent ions (Mg, Ca, Zn) it is into a M$_2$X$_4$ host, corresponding to a 1$e^-$ reduction per transition metal ion. \par

Since we consider discharge (or reduction reactions) at the cathode, a higher voltage implies a more thermodynamically favorable process. In the case of oxides with the LDP structure, intercalation is favored for most combinations of working ion and transition metal, in agreement with experimental and theoretical observations in monovalent and multivalent battery systems,{\cite{mizushima1980lixcoo2,thackeray1984electrochemical,arroyo2009gaining,yamada2006room,delmas1981,kim2015direct}} thus validating our approach. While certain combinations of monovalent cations, such as (Li/Na)-Cr, favor intercalation even in sulfides and selenides, most multivalent intercalation into $3d$-transition metal sulfides and selenides is expected to result in conversion. However, there may be a few compounds where intercalation into the LDP structure is kinetically stabilized, such as Li discharge into layered-NiO$_2$ ($V_{\rm int}^{\rm LDP} - V_{\rm conv}^{\rm LDP} = -$0.02~V) and Mg discharge into spinel-Ti$_2$S$_4$ ($V_{\rm int}^{\rm LDP} - V_{\rm conv}^{\rm LDP} = -$0.2~V), both of which are predicted to convert within our framework but are known to exhibit reversible intercalation experimentally.{\cite{ohzuku1993electrochemistry,sun2016high}} 

In contrast to oxides with the LDP structure, multivalent intercalation into all $3d$-oxides with the LCP structure is expected to undergo conversion. While LCP oxides are normally preferred in multivalent systems due to better multivalent mobility,{\cite{rong2015materials}} they are more likely to undergo conversion reactions instead of reversible intercalation, as illustrated by the experimental observations on Mg insertion in K-$\alpha$MnO$_2$.{\cite{arthur2014understanding}} In the case of multivalent sulfides and selenides, the tendency to convert becomes stronger with the LCP than the LDP structure, as indicated by the stronger conversion preference (stronger red squares) for Ca reduction of Fe, Co and Ni sulfides and selenides in Figure~{\ref{fig:3b}} compared to Figure~{\ref{fig:3}}. The tendency to convert becomes higher when LCP structures are used even for monovalent working ions, as signified by conversion and intercalation being favored with Li discharge in rutile-VO$_2$ (LCP) and layered-VO$_2$ (LDP), respectively, in agreement with experimental observations.{\cite{picciotto1984_LiVO2,murphy1978_rutile}}

For certain compounds, such as NaCrO$_2$, the intercalation voltage using both layered (LDP) and rutile (LCP) is higher than the corresponding conversion voltage, indicating that Na reduction of CrO$_2$ will always favor intercalation, regardless of polymorph. Also, our calculated voltage for Na$_x$CrO$_2$ is in good agreement with what has been reported experimentally,\cite{bo2016layered} further validating our approach. Most compounds, however, favor intercalation only when the LDP structure is used, as indicated by the data in Figures~{\ref{fig:3}} and {\ref{fig:3b}}.\par

\subsection{Two-electron Reduction}

\begin{figure}[h!]
\begin{center}
  \includegraphics[width=\columnwidth]{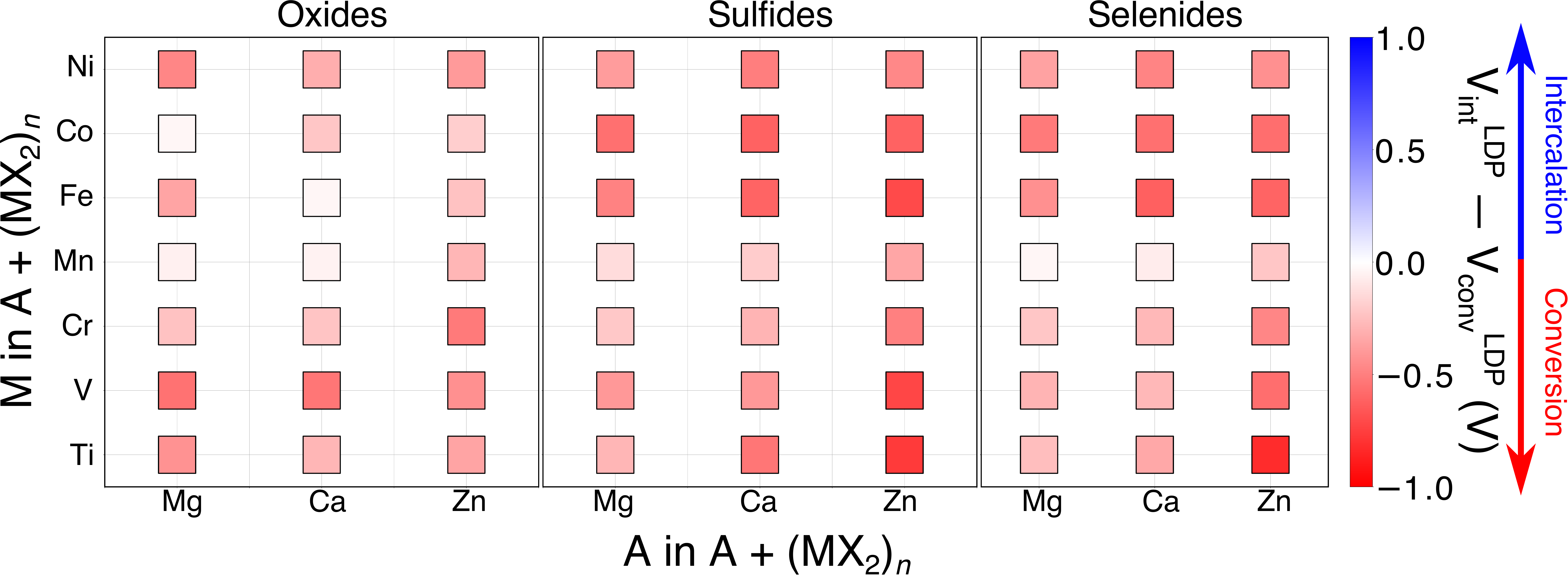}
  \end{center}
  \caption{\label{fig:4}
	Difference between the intercalation ($V_{\rm int}^{\rm LDP}$) and conversion ($V_{\rm conv}^{\rm LDP}$) voltage for 2-electron reduction reactions, starting from the lowest energy discharged polymorph (LDP). The voltage difference is indicated for three working ions (A~=~Mg, Ca, and Zn) in various $3d$-transition metal oxide, sulfide, and selenide hosts. Higher intercalation voltages are indicated by blue-colored squares while higher conversion voltages are red-colored. The intercalation of divalent ions considered is into a MX$_2$ structure, corresponding to a 2$e^-$ (per transition metal ion) reduction.}
\end{figure}

Figure~{\ref{fig:4}} plots the difference between the intercalation and conversion voltage for a 2$e^-$ reduction process in $3d$-transition metal oxide (left panel), sulfide (center) and selenide (right) hosts, using the structure of the lowest energy discharged polymorph (LDP). Here, two-electron reduction reactions are restricted to the MV working ions (Mg, Ca, and Zn). The potentially high capacities enabled by the transfer of two electrons per MV ion (and thus two electrons per redox center in the cathode) is one of the most appealing aspects of MV cathodes.\cite{canepa2017odyssey} 

In this case, however, the conversion voltages are higher for all compounds considered, indicating that intercalation is never favored for a two-electron reduction of the transition metal. While our analysis, which is restricted to compounds having an MX$_2$ stoichiometry and 3$d$ transition metals, does not rule out completely the possibility of high-capacity cathodes based on stoichiometric two-electron reduction, it does suggest that achieving two-electron reduction will require the exploration of other compositions. Similar to trends in the 1$e^-$ reduction process (Figures~{\ref{fig:3}} and {\ref{fig:3b}}), we expect conversion reactions to become more preferable when the lowest energy charged polymorph (LCP) is used for 2$e^-$ reduction (see Figure~S2 in SI).

\subsection{Thermodynamic Analysis}
\label{sec:thermo} 
\begin{figure}[h!]
\begin{center}
  \includegraphics[width=\columnwidth]{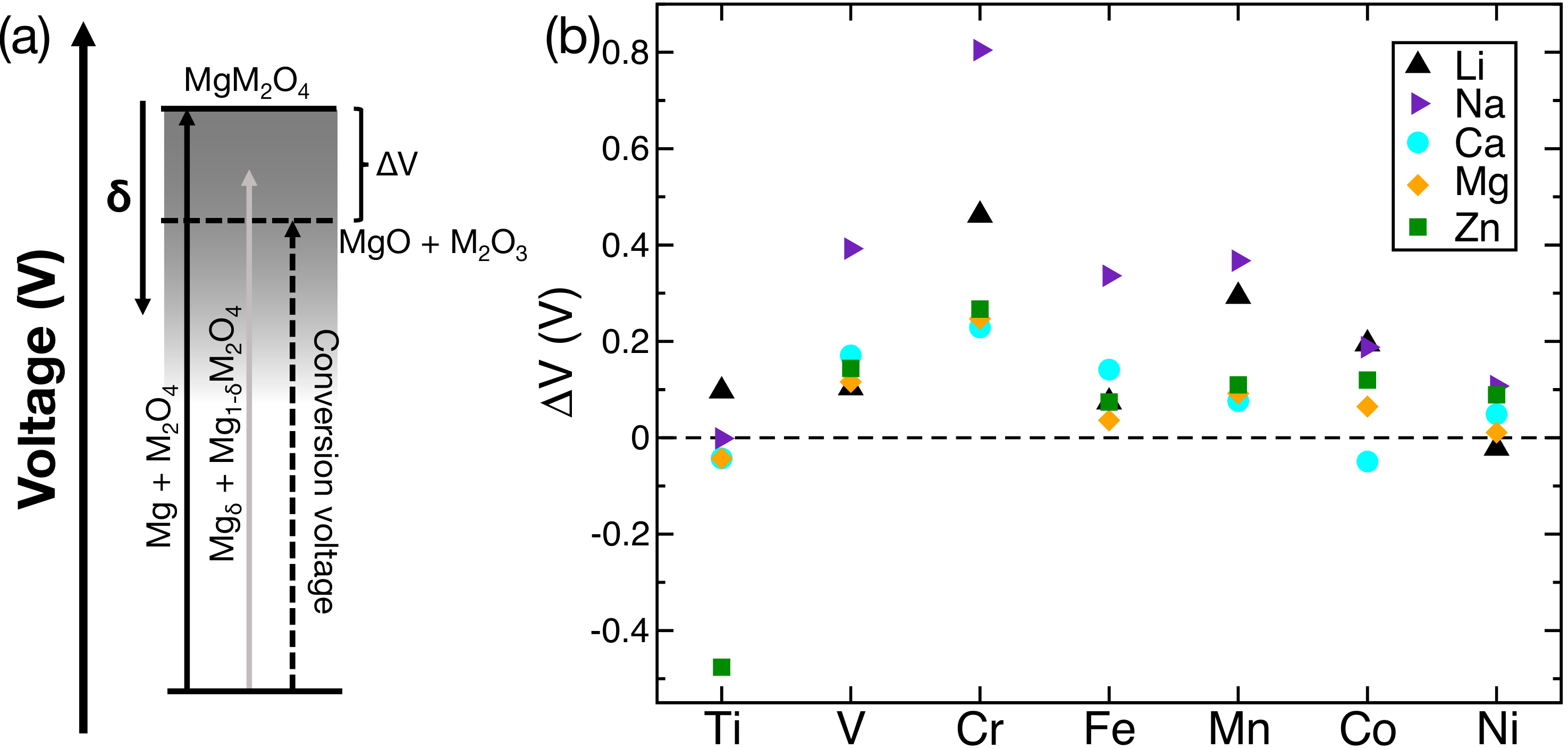}
  \end{center}
  \caption{\label{fig:5}
  (a) Schematic of resistance to conversion reactions. (b) The difference between intercalation and conversion voltages for each transition metal oxide for a one electron reduction. Specifically, $\Delta V = V_{\rm int}^{\rm LDP} -  V_{\rm conv}^{\rm LDP}$. Each colored point represents a working ion considered in this study, namely Li (black), Na (purple),  Ca (cyan), Mg (orange), and Zn (green).}
\end{figure}

As noted in the 1$e^-$ reduction reactions (Figures~{\ref{fig:3}} and {\ref{fig:3b}}), many compounds exhibit an energetic preference for intercalation or conversion depending on the polymorph that is being reduced. A higher potential for intercalation over conversion is needed to create stable intercalation cathodes. The magnitude of the difference between intercalation and conversion voltages, $\Delta V = V_{\rm int}^{\rm LDP} - V_{\rm conv}^{\rm LDP}$ (ideally $\Delta V > 0$), is important to keep the intercalation cathode stable even when poor kinetics induces large over/underpotentials. For example, Mg intercalation into a low mobility MV cathode can lead to accumulation of working ions in the surface region of the cathode material, causing the local voltage to drop to a value where conversion becomes thermodynamically possible (Figure~{\ref{fig:5}}a). Thus the difference in voltage between intercalation and conversion (labeled as $\Delta V$ on Figure \ref{fig:5}a)  directly determines the amount of local magnesiation inhomogeneity that can be tolerated. \par

Figure \ref{fig:5}b displays the estimated $\Delta V$ for each 3$d$ transition metal oxide for all five working ions, namely Li (black triangles), Na (purple triangles), Ca (cyan circles), Mg (orange diamonds), and Zn (green squares). The quantity $\Delta V$ is defined as $V_{\rm int}^{\rm LDP} - V_{\rm conv}^{\rm LDP}$. Note that $\Delta V$ is identical for both Mg/Ca discharge in Ti-oxides and for Li/Zn discharge in Fe-oxides, leading to an overlap of symbols in Figure~{\ref{fig:5}}. The data in Figure \ref{fig:5}b indicate that Cr oxides favor intercalation most strongly, while Ti-containing cathodes tend toward conversion.  The other transition metal oxides also favor intercalation (for several working ions) for one-electron reduction reactions with LDP structures, albeit not as strongly as Cr oxides. A similar analysis, presented in the Supporting Information of this work, finds that Cr sulfides and selenides also favor intercalation most strongly among the 3$d$ transition metals. We attribute the high resistance of Cr compounds toward conversion reactions to an ``ideal balance'' in the chemical potentials of Cr and the anion within the cathode host structure, as detailed in Section~S7 of SI. \\

\section{Discussion}

\subsection{Intercalation vs. Conversion: One-electron Reduction}

\subsubsection{The effect of anion chemistry}
Figure \ref{fig:2}a demonstrates the impact of anion variation on reaction voltage for the Mg-Cr-X system. As is apparent from the $V^{\rm MgX}_{\rm f}$ indicated on the $x$-axis of Figure \ref{fig:2}a, Mg reduction of Cr oxides, in the form of either intercalation or conversion, occurs at a higher voltage than with Cr sulfides or selenides, a trend that can be seen across different working ions and redox metals (Figure~{\ref{fig:3}}),{\cite{liu2015spinel,liu2016evaluation}} similar to what has been observed in Li-systems {\cite{aydinol1997ab}} The higher absolute voltage of oxides (than sulfides and selenides) in Li-systems is related to the higher electrostatic binding energy of the Li$^+$ ion and the lower energy levels of the transition metal in oxides.{\cite{aydinol1997ab}} The $3p$ orbitals of S and $4p$ of Se hybridize more extensively with the transition metals,{\cite{sharma1990_ionicity}} increasing the energy of these orbitals and thereby reducing the voltage.{\cite{seo2016_naturechem}} The better screening of the Li-anion interaction in sulfides or selenides and the larger volume further contribute to reduce the electrostatic energy gain{\cite{aydinol1997ab}} when Li$^+$ inserts into a sulfide or selenide host, further reducing the voltage. However, the comparatively stronger electrostatic interaction of MV ions with O$^{2-}$ than with S$^{2-}$ and Se$^{2-}$ can also be detrimental to battery performance, with oxides generally exhibiting lower MV ion mobility than the corresponding sulfides or selenides.\cite{liu2015spinel, liu2016evaluation, canepa2017high} \par

It is worth remarking that the voltage difference between various polymorphs in MgCr$_2$O$_4$ is much larger than in MgCr$_2$(S/Se)$_4$ (Figure~{\ref{fig:2}}a). In particular, the fact that $V^{\rm LDP}_{\rm int}$ is much higher than $V^{\rm LCP}_{\rm int}$ for MgCr$_2$O$_4$ partially reflects the substantial instability of the charged spinel Cr$_2$O$_4$ (E$^{\rm hull}$ = 194 meV/atom) compared to either the Cr$_2$S$_4$ or Cr$_2$Se$_4$ spinel (E$^{\rm hull}$ = 20 and 30 meV/atom, respectively). Thus, the ability of oxides to tolerate high levels of metastability,\cite{sun2016thermodynamic} particularly for charged-state structures, is critical for successful topotactic intercalation at high voltages.

A striking aspect of the data from 1$e^-$ reduction in LDP structures (Figure~{\ref{fig:3}}) is the overall preference of intercalation over conversion for most oxides with monovalent and multivalent ions, while only a few sulfides and selenides energetically prefer intercalation with multivalent working ions. Additionally, the tendency of LCP oxides to convert after 1$e^-$ reduction (Figure~{\ref{fig:3b}}) for all working ions is also significantly lower compared to sulfides and selenides. From the data in Figures~\ref{fig:3} and {\ref{fig:3b}}, we conclude that oxides generally favor intercalation reactions (except Ti-containing oxides), while sulfides and selenides are thermodynamically more likely to undergo conversion reactions (except Cr- and Mn-sulfides/selenides). In particular, (Li/Na)Cr(S/Se)$_2$, (Li/Na)Mn(S/Se)$_2$, and (Mg/Zn)Cr$_2$S$_4$ may potentially exhibit thermodynamically favorable intercalation reactions (Figure~{\ref{fig:3}}).\par

For systems such as Na$_{\rm x}$TiO$_2$, which exhibit $V_{\rm int}^{\rm LCP} < V_{\rm conv}^{\rm LCP}$ for one electron reduction (x=1, Figure~{\ref{fig:3b}}) conversion reactions are always favored at high degrees of reduction in the LCP (i.e., at x$\rightarrow$1). However, at x$<$1, TiO$_2$ can exhibit thermodynamically favorable Na intercalation if the system forms stable phases at any intermediate compositions, e.g.  Na$_{0.46}$TiO$_2$.{\cite{maazaz1983}} Indeed, DFT-calculated Na-Ti-O ternary phase diagram{\cite{jain2013commentary}} indicates the presence of several stable compositions for Na$_{\rm x}$TiO$_2$ (x$<$1), in agreement with experimental observations of reversible intercalation of less than one formula unit of Na per TiO$_2$ formula unit.{\cite{wang2015_tio2}}

Our analysis does not consider potential side-reactions that can occur in the presence of an electrolyte, either at the cathode/electrolyte interface{\cite{richards2016}} or within the cathode bulk.{\cite{sai2016_nano,kim2015_spinel_layered}} However, the data presented in this work can be taken as a guideline in identifying cathode chemistries that will be prone to conversion reactions \emph{irrespective} of the electrolyte used. For example, if the LDP for a given cathode chemistry is thermodynamically unstable, such as CaMn$_2$S$_4$ (Figure~{\ref{fig:3}}), Ca-discharge into MnS$_2$ will tend to form conversion products irrespective of the electrolyte or the polymorph of MnS$_2$ used.  While testing new cathode frameworks, especially in MV systems, robust characterization techniques must be used to verify that the electrochemical response observed is indeed intercalation instead of conversion.{\cite{canepa2017odyssey}}

\subsubsection{The effect of polymorph variation -- charged vs.~discharged states}

For battery cathodes undergoing electrochemical discharge, the process with the higher voltage (intercalation or conversion) will always drive the type of reduction process. Consequently, in the case of Mg discharge into CrO$_2$ (Figure~{\ref{fig:2}}b), we expect the rutile-CrO$_2$ polymorph to undergo conversion into MgO and Cr$_2$O$_3$ (dashed red arrow) while the spinel-CrO$_2$ polymorph (solid blue arrow) should yield the intercalated spinel-MgCr$_2$O$_4$ (Table~{\ref{table:new}}). Our findings suggest that polymorphism plays a crucial role in controlling the favored process between intercalation and conversion. 

The data in Figure \ref{fig:2}b (and results in Figures~{\ref{fig:3}} and {\ref{fig:3b}}) represent an important choice in the design of battery cathodes, i.e., whether to synthesize a cathode in the charged or discharged state. For example, in the Mg-Cr-O system (Figure~{\ref{fig:2}}b), Mg discharge into rutile-CrO$_2$ (which yields conversion) would presumably result from the preparation of the cathode in its stable charged state, while synthesis of the stable intercalated state, spinel-MgCr$_2$O$_4$, can potentially lead to electrochemically reversible Mg (de)intercalation. Although the empty spinel Cr$_2$O$_4$ is rather unstable ($E^{\rm hull} = 194$meV/atom), previous experimental studies of battery cathodes (such as spinel Mn$_2$O$_4$) indicate that it is possible to attain thermodynamically high energy charged-state structures following electrochemical extraction of the working ion after the cathode is synthesized in the intercalated state.\cite{thackeray1984electrochemical, arroyo2009gaining, zhou2006configurational, barpanda2013na2fep2o7} Therefore, synthesis of cathode materials in their stable intercalated forms can lead to better resistance against conversion reactions. \par 

An additional disadvantage of synthesizing a cathode in its stable charged state is the lower intercalation voltage, as indicated by lower $V_{\rm int}^{\rm LCP}$ than $V_{\rm int}^{\rm LDP}$ in Figure~{\ref{fig:2}}. Thermodynamically, a highly stable intercalated structure in combination with a highly metastable deintercalated structure always leads to higher intercalation voltages.{\cite{aydinol1997ab}} Also, electrochemical discharge into a stable charged state, in practice, will probably yield a metastable intercalated product, resulting in a voltage lower than $V_{\rm int}^{\rm LDP}$.\cite{mukherjee2017direct, sai2015intercalation} Thus, to achieve a higher voltage and a higher energy density, the synthesis of a cathode in its  intercalated state is preferable. Indeed, commercial Li-ion cathodes are always synthesized in their corresponding Li-intercalated frameworks.\cite{nitta2015li,whittingham2004}\par

While preparing a cathode in its stable intercalated form can potentially exhibit a higher intercalation voltage and superior resistance to conversion (see Figures~{\ref{fig:2}}, {\ref{fig:3}} and {\ref{fig:3b}}), sufficient working ion mobility has to be ensured in the cathode framework, especially in MV systems, thus highlighting the contradicting tradeoffs involved in emphasizing the stability of one end member or another (charged or discharged). Recent theoretical work has demonstrated that MV mobility can be enhanced by utilizing anion frameworks which host the MV ion in a ``less preferred'' coordination environment.\cite{rong2015materials} The preferred coordination environment can be determined for each working ion on a statistical basis considering known compounds containing the ion.{\cite{brown1988factors}} As most naturally occurring MV-compounds will host the MV ion in an preferred environment,{\cite{brown1988factors}} structures that \emph{do not} naturally contain MV ions are more likely to exhibit fast MV diffusion.{\cite{rong2015materials} Hence, synthesizing cathode frameworks in their stable charged-states increases the possibility of forcing the MV ion into a less preferred environment and consequently enhancing the MV mobility at the expense of voltage and conversion resistance.\par

While we consider only topotactic intercalation reactions, the thermodynamic framework used in our work is inclusive of the aspect of phase transitions during intercalation. For example, if Mg intercalation into rutile-CrO$_2$ (LCP, Figure~{\ref{fig:2}}) results (hypothetically) in spinel-MgCr$_2$O$_4$ (LDP), the voltage of the intercalation process, $V_{\rm int}^{\rm LCP \rightarrow LDP}$, includes a rutile~$\rightarrow$~spinel phase transition. Note that $V_{\rm int}^{\rm LCP \rightarrow LDP}$ is lower (higher) than the topotactic $V_{\rm int}^{\rm LDP}$ ($V_{\rm int}^{\rm LCP}$) since the charged (discharged) polymorph exhibits a lower Gibbs energy due to the phase transition (see Equation~{\ref{eq:1}}), i.e., $V_{\rm int}^{\rm LCP} \leq$ $V_{\rm int}^{\rm LCP \rightarrow LDP} \leq V_{\rm int}^{\rm LDP}$. Thus, the topotactic $V_{\rm int}^{\rm LCP}$ and $V_{\rm int}^{\rm LDP}$ signify ``reasonable'' lower and upper bounds of intercalation voltage, respectively, for a cathode chemistry. Any intercalation that includes a phase transition will likely exhibit a voltage between the two bounds.

Given the conflicting trade-offs that exist between preparing MV cathodes in their stable charged states (high mobility, low intercalation voltage, low resistance to conversion) vs.~their stable intercalated states (high intercalation voltage, high resistance to conversion, low mobility), it is of paramount importance to ($i$) discover naturally occurring intercalated frameworks that host MV ions in a less preferred environment, ($ii$) develop procedures to synthesize metastable charged states that topotactically match with a stable intercalated state. The spinel family of compounds, specifically the oxides and sulfides, display significant promise for the development of Mg-cathodes since they host Mg in a less preferred tetrahedral environment and are thermodynamically stable.\cite{rong2015materials, kim2015direct, liu2015spinel, gautam2017} Alternatively, a metastable charged polymorph can potentially be attained following chemical or electrochemical extraction of a ``removable'' ion, such as Li, Na or Cu.{\cite{sun2016high,aurbach2000prototype}} Subsequently, the metastable charged polymorph can reversibly intercalate MV ions and form stable intercalation products. Indeed, this strategy has been employed to attain the cathode materials for all of the fully-functional Mg batteries to date, namely, the Chevrel-Mo$_6$S$_8$ and spinel-Ti$_2$S$_4$,\cite{aurbach2000prototype, sun2016high} and may be the most promising path to a MV cathode that exhibits both high-voltage and high-mobility.\par

\subsection{Intercalation vs. Conversion: Two-electron Reduction}

The ability of transition metals to withstand two-electron reduction (per redox center) during intercalation is essential for enabling high-capacity cathodes. If the transition metal in an MV cathode can only tolerate one-electron reduction, then limitations in the number of available redox sites guarantees that these cathodes can furnish at most a capacity equivalent to stoichiometric intercalation of a monovalent working ion. Indeed, Figure \ref{fig:4} demonstrates that two-electron reduction of the transition metal in ternary chalcogenide frameworks always favors conversion, suggesting that for most compositions at the A(MX$_2$)$_n$ stoichiometry, MV intercala}tion cathodes exhibiting twice the capacity of their monovalent counterparts is unlikely. Of course the lack of high capacity cathodes does not discount the substantial gain in energy density resulting from the facile use of a metallic anode, a key advantage of MV batteries.\cite{canepa2017odyssey}\par

The data in Figures \ref{fig:3}, \ref{fig:4} and {\ref{fig:5}} underscore another major obstacle to MV batteries, namely the potential accumulation of MV ions near cathode particle surfaces due to low MV-ion mobility. Because two-electron reduction of the transition metal in A(MX$_2$)$_n$ compounds never favors intercalation (Figure \ref{fig:4}) and most compounds exhibit  a small degree of conversion resistance in the 1-electron limit (Figure \ref{fig:3}), the local accumulation of working ions could present a major issue. Indeed, Mg discharge into K-$\alpha$MnO$_2$ has been reported to produce MnO (implying a +2 Mn oxidation state) as a conversion product even at low degrees of discharge (187 mAh/g, corresponding to Mg$_{0.32}$MnO$_2$),{\cite{arthur2014understanding}} suggesting that enough Mg accumulated near the surface to yield a two-electron reduction, resulting in conversion. The thermodynamic difficulties associated with two-electron reduction highlights the importance of discovering cathode materials which permit fast diffusion of MV ions.\par

Finally, it is worth noting that in our prior work,\cite{canepa2017odyssey} we carried out a more limited form of this analysis and found that an overall two-electron intercalation with MV ions should in fact be possible in V$_2$O$_5$ and MoO$_3$, consistent with experimental results.\cite{gershinsky2013electrochemical} The Chevrel-phase Mo$_6$S$_8$ cathode is also capable of reversible two-electron reduction (per Mo$_6$S$_8$ f.u.), likely as a result of utilizing $4d$-Mo (which is stable in a large number of oxidation states). While cathodes like V$_2$O$_5$ and MoO$_3$ are less energy-dense in the limit of 1-electron reduction owing to a higher ratio of anions to redox centers, if two-electron reduction can be achieved, the capacity gains may be worthwhile. Another potential pathway to realize multi-electron redox (per f.u.) will be to employ polyanion systems.{\cite{canepa2017odyssey}} However, polyanion systems can also be susceptible to conversion reactions since the chemical space available for decomposition is higher (quaternary/quinary systems compared to A-M-X ternaries considered in this work), as indicated by conversion reactions during Mg discharge in FePO$_4$.{\cite{zhang2016unveil}} In summary, to design high capacity MV cathodes, either frameworks that are significantly different from the conventional A(MX$_2$)$_n$ compounds (e.g., V$_2$O$_5$\cite{mukherjee2017direct}, WO$_3${\cite{wang2017_wo3}} and VOPO$_4$\cite{wangoh2016uniform}), or materials that contain $4d$-transition metals with a large range of stable oxidation states (such as Mo-containing MoO$_3$ or RuO$_2$\cite{gregory1990nonaqueous}), may have to be sought after.

\subsection{The magnitude of intercalation (or conversion) preference}

\subsubsection{Kinetic stabilization of metastable phases}
It is important to note that the data in Figures \ref{fig:3}, {\ref{fig:3b}} and \ref{fig:4} reflect thermodynamic quantities not accounting for the sometimes-crucial role played by the kinetic stabilization of metastable phases. For example, a functional Mg intercalation battery was constructed using a spinel Ti$_2$S$_4$ cathode,{\cite{sun2016high}} a compound we predict will favor conversion ($\Delta V = -0.2$~V, Figure \ref{fig:3}b), which highlights the possibility for kinetic stabilization to enable intercalation reactions even when conversion is thermodynamically favored. While the energetic limits of kinetic stabilization in this context are not rigorously known and may be chemistry-specific, we may empirically consider $\Delta V = -0.2$~V as a cutoff below which kinetic stabilization might occur, given the experimental evidence for some compounds that function in this range. In this context, compounds for which conversion occurs at only a slightly higher voltage than intercalation ($\Delta V > -$0.2~V) remain plausible candidates for further investigation. Examples in Figure~{\ref{fig:3}} include MgCr$_2$Se$_4$, CaCo$_2$O$_4$, and ZnCo$_2$S$_4$. Larger separation between the conversion and intercalation voltages (i.e., $\Delta V < -$0.2~V) means that kinetically-stabilized intercalation is less likely to occur. As a benchmark, $V_{\rm int}^{\rm LCP} - V_{\rm conv}^{\rm LCP} = -$0.4~V for $\alpha$-MnO$_2$, which is known experimentally to convert.\cite{arthur2014understanding}\par

Extending our analysis to compounds in Figure~{\ref{fig:4}}, we found a few compositions that exhibit two-electron intercalation voltages less than 0.2 V below the conversion voltages (i.e., $\Delta V > -0.2$~V): MgCoO$_2$, CaMnO$_2$, MgMnSe$_2$, and CaMnSe$_2$. These chemistries may exhibit reversible intercalation and are interesting candidates for further investigation in the context of high-capacity cathodes.{\cite{okamoto2015}}

\subsubsection{Resistance to conversion reactions}
A large, positive $\Delta V$ (a measure of conversion resistance, as explained in Section \ref{sec:thermo} and demonstrated in Figure \ref{fig:5}a) is especially critical for low-mobility MV ions and rapid rates of discharge, where substantial local accumulation of the working ion can occur. As shown in Figure \ref{fig:5}b for oxides, $\Delta V$ first increases then decreases as the transition metal varies across the 3d period, peaking at Cr. Thus, Cr-containing compounds are expected to show the highest resistance to conversion reactions for both monovalent and multivalent working ions.  The tendency of Cr-compounds to resist conversion could arise due to an ``optimal" combination of Cr and anion chemical potentials, as discussed in the Section~S7 of SI.

In the case of sulfides and selenides, there are no clear trends in the variation of the transition metal or anion chemical potential across the $3d$-series (see Figures~S4 and S5 in the SI), despite Cr-compounds exhibiting the highest $\Delta V$ (Figure~S3). However, we can infer from inspection of Figure~\ref{fig:3} (and Figure~{\ref{fig:3b}}) that most of the sulfide and selenide compounds which favor intercalation contain Cr (Mg, Zn, Na, and Li intercalation into CrS$_2$ and CrSe$_2$). Thus, we can conclude that Cr-containing chalcogenides are the most resistant to conversion among all $3d$-metals with an A(MX$_2$)$_n$ framework.\par

\section{Conclusion}
In this work we have studied the thermodynamics of intercalation and conversion reactions in A(MX$_2$)$_n$ (A = Na, Li, Mg, Ca, Zn; M = 3$d$ transition metal, X = O, S, Se) compounds. We demonstrated the important tradeoffs involved in selecting an anion chemistry. While prior work has demonstrated that sulfides and selenides generally exhibit higher mobility,\cite{liu2016evaluation, canepa2017high} we have shown here that they react at lower voltages and have higher driving forces for conversion than oxides. Polymorph selection was shown here to be similarly important: While cathode polymorphs stable in the charged state are more likely to exhibit favorable MV ion mobility, they necessarily react at lower voltages and are less resistant to conversion reactions. Our findings suggest that certain polymorphs of a given chemistry favor conversion (rutile CrO$_2$) while others favor intercalation (spinel CrO$_2$).\par 

We also find that two-electron reduction of the transition metal will always favor conversion reactions for A(MX$_2$)$_n$ compounds.  While the A(MX$_2$)$_n$ stoichiometry is common in battery cathode materials, there are other possible stoichiometries which have not been considered in the present analysis and should be investigated. In addition to reporting conversion and intercalation reaction voltages across structural and chemical space and discussing the resulting implications, we also identify a remarkable ability of Cr-based cathode materials to resist conversion reactions.  

Overall, we conclude that Cr offers the most promising choice of transition metal and that $3d$-oxides provide superior voltage and conversion resistance. Taking these results into account, we conclude that the simultaneous requirement for high voltage and high mobility for MV ions can be best met by seeking out materials which contain an extractable ion in an environment that is non-preferred for the MV working ion. This is consistent with the identification of the two most reversible Mg cathodes so far, TiS$_2${\cite{sun2016high}} and Mo$_6$S$_8$.{\cite{aurbach2000prototype}} Thus, in the context of MV cathode design, high voltage and conversion resistance have to be carefully balanced with sufficient MV mobility.
\section*{Acknowledgment}
%
The current work is fully supported by the Joint Center for Energy Storage Research, an Energy Innovation Hub funded by the U.S. Department of Energy (DOE), Office of Science and Basic Energy Sciences. This study was supported by Subcontract 3F-31144. The authors thank the National Energy Research Scientific Computing Center (NERSC) for computing resources.
\bibliography{library} 
%

\end{document}